\documentclass[prd,aps,amsmath,showpacs,nofootinbib]{revtex4}

\usepackage{times}
\usepackage{hyperref}

\usepackage{graphicx}
\usepackage{dcolumn}
\usepackage{bm}
\usepackage{color} 

\begin{document}

\title{Radiative corrections to the Dalitz plot of $K_{l3}^0$ decays}

\author{
M.\ Neri
}
\affiliation{
Departamento de F{\'\i}sica, Escuela Superior de F\'{\i}sica y Matem\'aticas del IPN, Apartado Postal 75-702, M\'exico, D.F.\ 07738, Mexico
}

\author{
A.\ Mart{\'\i}nez
}
\affiliation{
Departamento de F{\'\i}sica, Escuela Superior de F\'{\i}sica y Matem\'aticas del IPN, Apartado Postal 75-702, M\'exico, D.F.\ 07738, Mexico
}

\author{
C.\ Ju\'arez-Le\'on
}
\affiliation{
Departamento de F{\'\i}sica, Escuela Superior de F\'{\i}sica y Matem\'aticas del IPN, Apartado Postal 75-702, M\'exico, D.F.\ 07738, Mexico
}

\author{
J.\ J.\ Torres
}
\affiliation{
Departamento de Posgrado, Escuela Superior de C\'omputo del IPN, Apartado Postal 75-702, M\'exico, D.F.\ 07738, Mexico
}

\author{
Rub\'en Flores-Mendieta
}
\affiliation{
Instituto de F{\'\i}sica, Universidad Aut\'onoma de San Luis Potos{\'\i}, \'Alvaro Obreg\'on 64, Zona Centro, San Luis Potos{\'\i}, San Luis Potos{\'\i} 78000, Mexico
}

\date{\today}

\begin{abstract}
A model-independent expression for the Dalitz plot of semileptonic decays of neutral kaons, $K_{l3}^0$, including radiative corrections to order $(\alpha/\pi)(q/M_1)$, where $q$ is the momentum transfer and $M_1$ is the mass of the kaon, is presented. The model dependence of radiative corrections is kept in a general form within this approximation, which is suitable for model-independent experimental analyses. Expressions for bremsstrahlung radiative corrections are presented in two forms: one with the triple integral over the kinematical variables of the photon ready to be performed numerically and the other one in a fully analytical form. The final result is restricted to the so-called three-body region of the Dalitz plot and it is not compromised to fixing the values of the form factors at predetermined values.
\end{abstract}

\pacs{14.40.Df, 13.20.Eb, 13.40.Ks}

\maketitle

\section{\label{sec:intro}Introduction}

Nowadays it is well established in the standard model that the transitions between weak charged currents mix quarks of different generations, which is encoded in the Cabibbo-Kobayashi-Maskawa (CKM) matrix. Over the years, a substantial amount of effort of both the experimental and theoretical bent has gone into the determination of the elements of this matrix. The most precise constraints on the size of these matrix elements are extracted from the low-energy $s\to u$ and $d\to u$ semileptonic transitions; therefore, $V_{us}$ and $V_{ud}$ possess particular interest because unitarity can be better tested in the first row of the CKM matrix so that the validity of the relation
\begin{equation}
|V_{ud}|^2+|V_{us}|^2+|V_{ub}|^2 =1
\end{equation}
can be probed at the $0.1\%$ level \cite{ckm05}.

The most precise determination of $|V_{ud}|$ comes from the analysis of superallowed $0^+\to 0^+$ nuclear beta decays, whereas $|V_{us}|$ can be better determined from the semileptonic decays of $K$ mesons ($K_{l3}$ decays) and (to a minor extent) hyperons and also from $\tau$ decays.

The decay rates of all $K_{l3}$ modes ($K = K^\pm, K^0$, $l=e,\mu$) can be expressed as \cite{ckm05}
\begin{equation}
\Gamma(K_{l3[\gamma]}) = \frac{G_F^2 S_{\mathrm{ew}} M_1^5}{128 \pi^3} C_K^2 I^{Kl}(\lambda_i) |V_{us} f_+^{K^0\pi^-}(0)|^2 \left[1 + 2 \Delta_{SU(2)}^K + 2\Delta_{\mathrm{EM}}^{Kl} \right], \label{eq:dgamma}
\end{equation}
where $G_F$ is the Fermi constant, $C_K$ is a Clebsh-Gordan coefficient that is equal to 1 and $1/\sqrt{2}$ for $K_{l3}^0$ and $K_{l3}^\pm$ decays, respectively, and $M_1$ is the mass of the decaying kaon; for ease of notation, $G=C_K V_{us} G_F$ hereafter. Additionally, $S_{\mathrm{ew}}$ comprises the short distance electroweak correction to semileptonic charged-current processes, and $I^{Kl}(\lambda_i)$ is a phase-space integral that depends on the slope and curvature of the form factor $f_+^{K^0\pi^-}(0)$, which as indicated in Eq.~(\ref{eq:dgamma}) is customarily used to normalize the form factors of all channels. Actually, for the experimental extraction of $|V_{us}f_+(0)|$ the neutral decay $K_L^0\to \pi e \nu$ is preferred in order to avoid isospin-symmetry breaking corrections that appear in $K_{l3}^\pm$ decays and the complications introduced by an additional scalar form factor in $K_{\mu 3}$ decays. Finally, the terms $\Delta_{SU(2)}^K$ and $\Delta_{\mathrm{EM}}^{Kl}$, which are channel dependent, represent the isospin-breaking and long distance electromagnetic corrections, respectively. A determination of $|V_{us}|$ at the $1\%$ level requires the inclusion of both corrections.

The main aim of the this paper is precisely the computation of the radiative corrections (RC) to differential decay rate ---or equivalently, the Dalitz plot (DP)---of $K_{l3}^0$ decays, following the approach implemented in the analogous analysis for the charged counterpart, $K_{l3}^\pm$ decays, presented in previous works \cite{jl2011,jl2012}. The approach leads to an analytical expression that comprises contributions of both virtual and real photons, restricted to the three-body part of the allowed kinematical region, hereafter referred to as the three-body region (TBR).

There are various works addressing the radiative corrections to $K_{l3}$ decays, each one from a different perspective. An important selection of such analyses are those by Ginsberg \cite{ginsKl3pm,ginsKe3pmDP,ginsKe30,ginsKm3}, Becherrawy \cite{bech}, Garc{\'\i}a and Maya \cite{maya}, Cirigliano \textit{et.\ al.} \cite{cir,cir2}, Bytev \textit{et al.} \cite{bytev}, and Andre \cite{andre}. Ginsberg calculated the radiative corrections to the lepton spectrum, DP, and decay rates of $K_{l3}$ decays, by assuming a phenomenological weak $K$-$\pi$ vertex. Becherrawy used a particular model of the strong interactions. Garc{\'\i}a and Maya extended to $M_{l3}$ Sirlin's approach \cite{sirlin}, originally introduced to study the radiative corrections to the charged lepton spectrum in neutron beta decay. Cirigliano \textit{et al.} implemented chiral perturbation theory and accounted for virtual photons and leptons. Bytev \textit{et al.} removed the ultraviolet cutoff dependence by setting it equal to the $W$ mass. Andre included contributions from outside the kinematically allowed TBR of the DP in $K_{l3}^0$ decays.

It is quite hard \textit{a priori} to assess the success of the different approaches in the calculation of RC to $K_{l3}$ decays. This paper was written in response to the need of having a reliable expression that is free from an infrared divergence, that does not contain an untraviolet cutoff, and above all, that is not compromised by any model dependence of RC. The above criteria are met by the final expression presented here so its applicability to model-independent Monte Carlo analyses is immediate.

The ordering of the paper is as follows. Section \ref{sec:virt} is mostly devoted to introduce the notation and conventions used through the basics on kaon semileptonic decays. The calculation of virtual RC to order $(\alpha/\pi)(q/M_1)$ is also presented. Section \ref{sec:bre} is intended to provide results of the bremsstrahlung RC in the triple numerical integration form and combine them with the virtual RC part to obtain the first main result. In addition, the corresponding fully analytical expressions are also given, which yield to the second main result. The last section, \ref{sec:finr}, is dedicated to a summary and to concluding remarks. For completeness, a comparison with other results available in the literature is also performed. The comparison is satisfactory.

\section{\label{sec:virt}Virtual radiative corrections}

In this section the notation and conventions are first introduced and afterwards the virtual radiative corrections are calculated.e

The four-momenta and masses of the particles involved in the semileptonic decay of a neutral kaon
\begin{equation}
K^0(p_1) \to \pi^-(p_2) + \ell^+(l) + \nu_\ell (p_\nu^0), \label{eq:kl3}
\end{equation}
will be denoted by $p_1=(E_1,{\mathbf p}_1)$, $p_2=(E_2,{\mathbf p}_2)$, $l=(E,{\mathbf l})$, and $p_\nu=(E_\nu^0,{\mathbf p}_\nu^0)$, and by $M_1$, $M_2$, $m$, and $m_\nu$, respectively. No assumptions will be made about the size of $m$ compared to $M_1$ so the final expressions obtained will be valid for both $K_{e3}^0$ and $K_{\mu 3}^0$ decays alike. When the calculation is specialized to the center-of-mass frame of the decaying kaon, $p_2$, $l$, and $p_\nu$ will also denote the magnitudes of the corresponding three-momenta. Also, the direction of a generic three-vector ${\mathbf p}$ will be denoted by a unit vector $\hat {\mathbf p}$.

The uncorrected transition amplitude $\mathsf{M}_0$ (i.e., the amplitude without RC) for process~(\ref{eq:kl3}) can be readily obtained by keeping only the contribution of the vector current and neglecting scalar and tensor contributions. $\mathsf{M}_0$ can thus be written as \cite{jl2011}
\begin{eqnarray}
\mathsf{M}_0 = \frac{G}{\sqrt 2} W_\mu(p_1,p_2) \left[\overline{u}_\nu(p_\nu) O_\mu v_\ell(l)\right], \label{eq:M0}
\end{eqnarray}
where $v_\ell$ and $u_\nu$ are the Dirac spinors of the corresponding particles, $O_\mu \equiv \gamma_\mu(1+\gamma_5)$, and the metric and $\gamma$-matrix convention adopted here is the standard one (see, for instance, Ref.~\cite{bd}), except that $\gamma_5$ has the opposite sign. The hadronic matrix element $W_\mu(p_1,p_2)$ is given by
\begin{eqnarray}
W_\mu (p_1,p_2) & = & \langle \pi^-(p_2)|\bar{u} \gamma_\mu s|K^0(p_1) \rangle \nonumber \\
& = & f_+(q^2) (p_1+p_2)_\mu + f_-(q^2) (p_1-p_2)_\mu, \label{eq:mtxel}
\end{eqnarray}
where $q=p_1-p_2$ is the four-momentum transfer and $f_\pm (q^2)$ are dimensionless form factors.

Armed with the transition amplitude $\mathsf{M}_0$, the differential decay rate in the variables $E$ and $E_2$ (Dalitz plot) can be obtained straightforwardly. It can be expressed as
\begin{equation}
d\Gamma_0(K_{l3}) = A_0 d\Omega,
\end{equation}
where
\begin{equation}
A_0 = A_1^{(0)} |f_+(q^2)|^2 + A_2^{(0)} \mathrm{Re} [f_+(q^2)f_-(q^2)] + A_3^{(0)} |f_-(q^2)|^2. \label{eq:a0}
\end{equation}
The functions $A_i^{(0)}$ are given in Eqs.~(17)--(19) of Ref.~\cite{jl2011} and the factor $d\Omega$ reads
\begin{equation}
d\Omega = C_K^2 G_F^2 \frac{|V_{us}|^2}{32\pi^3} M_1^3 dEdE_2.
\end{equation}

The method to calculate the virtual RC to the DP of $K_{l3}^\pm$ decays has been discussed in detail in Ref.~\cite{jl2011}. It can be readily adapted to the present case, so only a few salient facts will be repeated now.

To first order in $\alpha$, the Feynman diagrams which yield the virtual RC in $K_{l3}^0$ decays are similar to the ones depicted in Fig.~2 of Ref.~\cite{jl2011}. Basically they comprise graphs in which the virtual proton is emitted from the hadronic line or the intermediate vector boson and is absorbed by the charged lepton. The contribution reads,
\begin{equation}
\mathsf{M}_{V_1} = \frac{G}{\sqrt{2}} \frac{\alpha}{4\pi^3i} \int d^4k \left[\frac{D_{\mu\alpha} (k)}{k^2-2l\cdot k+i\epsilon}\right] \left[\frac{W_\lambda (p_1,p_2) (2p_2+k)_\mu}{k^2+2p_2 \cdot k+i\epsilon} + T_{\mu\lambda} (p_1,p_2,k)\right] \overline{u}_\nu O_\lambda (2l_\alpha - \not k \gamma_\alpha) v_l,
\end{equation}
where $k$ is the virtual-photon four-momentum and $D_{\mu\alpha}(k)$ is the photon propagator. The first summand in the above equation is independent of the details of the strong interaction; it is free of the ultraviolet divergence, but contains the infrared divergence. On the contrary, all the model dependence due to the effects of the strong interactions is contained in $T_{\mu\lambda}$. For the purposes of this paper, no further details are needed here. The complete material nevertheless can be found in the original paper \cite{sirlin}, which was further adapted to Ref.~\cite{jl2011}.

Next, the lepton wave function renormalization graph yields
\begin{eqnarray}
\mathsf{M}_{V_2} & = & \frac{G}{\sqrt{2}} \frac{\alpha}{8\pi^3i} W_\lambda (p_1,p_2) \int d^4k D_{\alpha\mu} (k)\overline{u}_\nu O_\lambda \frac{\not l-m}{2m^2} \frac{(2l_\alpha + \gamma_\alpha \! \not k) {\not l} (2l_\mu+\not k\gamma_\mu)}{(k^2+2l\cdot k+i\epsilon)^2} v_l.
\end{eqnarray}

Finally, the graphs in which the photon is emitted by a hadronic line or the intermediate vector boson and is absorbed by the same hadronic line or another one or the intermediate boson yield the contribution
\begin{eqnarray}
\mathsf{M}_{V_3} & = & \frac{G}{\sqrt{2}} \frac{\alpha}{8\pi^3i} W_\lambda (p_1,p_2) \overline{u}_\nu O_\lambda v_l \int d^{4}k D_{\mu\alpha}(k) \frac{(2p_2-k)_\mu (2p_2-k)_\alpha}{(k^2-2p_2 \cdot k+i\epsilon)^2}+\mathsf{M}_{v_3}^\prime \nonumber \\
&  & = \mathsf{M}_{V_3}^c + \mathsf{M}_{V_3}^\prime.
\end{eqnarray}

Following Ref.~\cite{jl2011}, the virtual RC can be separated into a model-independent part $\mathsf{M}_V^i$ that is finite and calculable and into a model-dependent one that contains the effects of the strong interactions and the intermediate vector boson. The latter is contained in the term proportional to $T_{\mu\lambda}$ of $\mathsf{M}_{V_1}$ and in $\mathsf{M}_{V_3}^\prime$ of $\mathsf{M}_{V_3}$. To order $(\alpha/\pi)(q/M_1)$, such a model dependence amounts to two form factors $a_1^{\prime\prime}(q^2,p_+\cdot l)$ and $a_2^{\prime\prime}(q^2,p_+\cdot l)$, with $p_+ = p_1+p_2$, which can be absorbed into $f_+$ and $f_-$ of $\mathsf{M}_0$, respectively,  through the definition of effective form factors, hereafter referred to as $f_+^\prime$ and $f_-^\prime$. 

After the $k$ integration, the decay amplitude with virtual RC, $\mathsf{M}_V$, is given by
\begin{equation}
\mathsf{M}_V = \mathsf{M}_0^\prime \left[1 + \frac{\alpha}{2\pi}\Phi_{n}\right] - \frac{\alpha}{2\pi}\mathsf{M}_{p_2} \Phi_{n}^\prime,
\end{equation}
where the amplitudes $\mathsf{M}_0^\prime$ and $\mathsf{M}_{p_2}$ read
\begin{equation}
\mathsf{M}_0^\prime = \frac{G}{\sqrt{2}} \left[f_+^\prime(q^2,p_+\cdot l)(p_1+p_2)_\alpha + f_-^\prime(q^2,p_+\cdot l)(p_1-p_2)_\alpha \right] \bar{u}_\nu(p_\nu) O_\alpha v_l(l), \label{eq:m0prime}
\end{equation}
and
\begin{equation}
\mathsf{M}_{p_2} = \frac{1}{m} \frac{G}{\sqrt{2}} W_{\alpha}(p_1,p_2) \bar{u}_\nu(p_\nu) O_\alpha {\not \! p}_2 v_l(l).
\end{equation}
The prime on $\mathsf{M}_0$ in Eq.~(\ref{eq:m0prime}) is used as a reminder that the effective form factors appear explicitly in this amplitude. The model-independent functions $\Phi_{n}(E,E_2)$ and $\Phi_n^\prime(E,E_2)$ can be written as \cite{tun93}
\begin{eqnarray}
\Phi_n(E,E_2) & = & 2\left[\frac{1}{\beta^\prime} \mathrm{arctanh}\, \beta^\prime-1\right] \ln \frac{\lambda}{m} + \frac{\pi^2}{\beta^\prime} - \frac{1}{\beta^\prime} (\mathrm{arctanh}\, \beta^\prime)^2 - \frac{11}{8} + \left[\frac32 - \frac{m^2}{(p_2+l)^2}\right] \ln \frac{M_2}{m} \nonumber \\
&  & \mbox{} + \frac{1}{\beta^\prime}\left[L\left(\frac{\delta}{x_2^+}\right) + L\left(\frac{\delta}{1-x_2^-}\right)\right] + \frac{1}{\beta^\prime} \ln\left(\frac{1-x_2^+}{1-x_2^-} \right) \left[\ln \frac{M_2}{m} - \mathrm{arctanh}\, \beta^\prime - \frac12 \ln \left(\frac{1-x_2^+}{1-x_2^-} \right) \right] \nonumber \\
&  & \mbox{} + \frac{1}{\beta^\prime} \mathrm{arctanh}\, \beta^\prime \left[\frac{M_2^2 + p_2\cdot l(1+{\beta^\prime}^2)}{(p_2+l)^2} \right] + \frac{i\pi}{\beta^\prime} \left[\ln\frac{(p_2+l)^2}{\lambda^2} + 2\ln\delta - \frac{M_2^2 + p_2 \cdot l(1+{\beta^\prime}^2)}{(p_2+l)^2} \right], \label{eq:phin}
\end{eqnarray}
and
\begin{equation}
\Phi_n^\prime(E,E_2) = -\frac{m^2}{\beta^\prime p_2 \cdot l(p_2+l)^2} \left[(M_2^2 + p_2 \cdot l) \mathrm{arctanh}\, \beta^\prime + \beta^\prime p_2\cdot l \ln \frac{M_2}{m} - i\pi (M_2^2 + p_2\cdot l) \right], \label{eq:phinp}
\end{equation}
where
\begin{equation}
\beta^\prime = \sqrt{1-\frac{M_2^2 m^2}{(p_2 \cdot l)^2}}, \label{eq:betap}
\end{equation}
\begin{equation}
x_2^\pm = \frac{m^2 + p_2\cdot l(1\pm\beta^\prime)}{(p_2+l)^2},
\end{equation}
and
\begin{equation}
\delta = x_2^+ - x_2^-. \label{eq:ddelta}
\end{equation}
Here $L(x)$ is the Spence function and $\lambda$ is the infrared-divergent cutoff. In Eqs.~(\ref{eq:phin})--(\ref{eq:ddelta}) $p_2$ and $l$ are understood to be four-vectors. The term $\pi^2/\beta^\prime$, usually referred to as the Coulomb term, is characteristic of the electric interaction between charged particles of processes such as (\ref{eq:kl3}). It is an important contribution to the RC. 

The differential decay rate with virtual RC can now be obtained by using standard techniques, namely, by squaring $\mathsf{M}_V$, summing over spins in the final state and introducing appropriate phase-space factors. To order $(\alpha/\pi)(q/M_1)$ the resultant expression is
\begin{equation}
d\Gamma_V = d\Omega \left[\left(1+\frac{\alpha}{\pi} \mathrm{Re}\, \Phi_n \right) A_0^\prime + \frac{\alpha}{\pi} \mathrm{Re}\, \Phi_n^\prime A_{Vn}^\prime \right], \label{eq:dgv}
\end{equation}
where $A_0^\prime$ is the same expression given by Eq.~(16) of Ref.~\cite{jl2011} and $A_{Vn}^\prime$ reads
\begin{equation}
A_{Vn}^\prime = A_{1n}^{(V)} |f_+(q^2,p_+ \cdot q)|^2 + A_{2n}^{(V)} \mathrm{Re} [f_+(q^2,p_+ \cdot q)f_-^*(q^2,p_+ \cdot q)] + A_{3n}^{(V)} |f_-(q^2,p_+ \cdot q)|^2, \label{eq:avn}
\end{equation}
where
\begin{equation}
A_{1n}^{(V)} = \frac{4}{M_1^2} \left[ (M_1E_2+M_2^2) \frac{E_\nu^0}{M_1} - (M_1^2-M_2^2) \frac{E_m-E}{2M_1} \right],
\end{equation}
\begin{equation}
A_{2n}^{(V)} = \frac{8}{M_1^2} \left[ E_2E_\nu^0 - (M_1^2+M_2^2) \frac{E_m-E}{2M_1}  \right],
\end{equation}
and
\begin{equation}
A_{3n}^{(V)} = \frac{4}{M_1^2} \left[ (M_1E_2-M_2^2) \frac{E_\nu^0}{M_1} - (M_1^2-M_2^2) \frac{E_m-E}{2M_1} \right],
\end{equation}
where $E_{m}$ is the maximum energy of the charged lepton given by
\begin{equation}
E_m = \frac{M_1^2-M_2^2+m^2}{2M_1}.
\end{equation}

The effective form factors $f_\pm(q^2,p_+ \cdot q)$ displayed explicitly in Eq.~(\ref{eq:avn}) contain energy-dependent contributions of the model dependence in the virtual radiative corrections.

Once the virtual RC to the DP of process (\ref{eq:kl3}) have been calculated, the corresponding correction when a real photon is involved can be analyzed. This is done in the next section.

\section{\label{sec:bre}Bremsstrahlung radiative corrections}

In addition to the virtual RC, the bremsstrahlung counterpart must be added to obtain the complete RC to the DP of process (\ref{eq:kl3}). For this purpose, the four-body decay
\begin{equation}
K^0(p_1) \to \pi^-(p_2) + \ell^+(l) + \nu_\ell (p_\nu) + \gamma(k), \label{eq:kl3f}
\end{equation}
must be considered, restricted to the TBR of the allowed kinematical region \cite{jl2011}. In Eq.~(\ref{eq:kl3f}), $\gamma$ represents a photon with four-momentum $k=(\omega,\mathbf{k})$. Energy and momentum conservation yield $E_1=E_2+E+E_\nu+\omega$ and $\mathbf{p_1}=\mathbf{p_2}+\mathbf{l}+\mathbf{p}_\nu+\mathbf{k}$.

The Low theorem \cite{low,chew} will be used to obtain the bremsstrahlung amplitude $\mathsf{M}_B$ with all the $(\alpha/\pi)(q/M_1)$ contributions. The theorem states that the radiative amplitudes of orders $\mathcal{O}(1/k)$ and $\mathcal{O}(k^0)$ can be determined in terms of the nonradiative amplitude without further structure dependence.

The Feynman diagrams that yield $\mathsf{M}_B$ can be worked out in parallel with the ones depicted in Fig.~3 of Ref.~\cite{jl2011}. Skipping details, $\mathsf{M}_B$ can be written as
\begin{equation}
\mathsf{M}_B = \mathsf{M}_{B_1}+\mathsf{M}_{B_2}+\mathsf{M}_{B_3}+\mathsf{M}_{B_4}, \label{eq:MB}
\end{equation}
where the different contributions read
\begin{equation}
\mathsf{M}_{B_1} = -e\mathsf{M}_0 \left[\frac{l\cdot\epsilon}{l\cdot k} - \frac{p_2 \cdot\epsilon}{p_2 \cdot k} \right],
\end{equation}
\begin{equation}
\mathsf{M}_{B_2} = -\frac{eG}{\sqrt{2}} W_{\lambda} \bar{u}_{\nu}O^{\lambda} \frac{{\not k} \! {\not\epsilon}}{2l\cdot k} v_l,
\end{equation}
\begin{equation}
\mathsf{M}_{B_3} = -\frac{eG}{\sqrt{2}} (f_+-f_-) \left[-\frac{p_2 \cdot\epsilon}{p_2 \cdot k} k_{\lambda} + \epsilon_{\lambda}\right] L^{\lambda},
\end{equation}
and
\begin{equation}
\mathsf{M}_{B_4} = -\frac{eG}{\sqrt{2}} \left[\frac{\partial}{\partial q^2}(W_{\lambda} L^{\lambda}) \right] \left[2p_2 \cdot\epsilon\frac{p_1\cdot k}{p_2 \cdot k} - 2 p_1 \cdot\epsilon\right],
\end{equation}
where $L^\lambda = \bar{u}_\nu O^\lambda v_l$.
Observe that (\ref{eq:MB}) is gauge invariant and model independent. 

While the amplitude $\mathsf{M}_{B_1}$ contains terms of order $\mathcal{O}(1/k)$, the other ones contain terms of order $\mathcal{O}(k^0)$. Furthermore, the contribution of $\mathsf{M}_{B_4}$ will be neglected because it yields terms of order $q^2/M_1^2$ to the decay rate, which are not needed in the present analysis. The infrared-divergent terms are thus all contained in $\mathsf{M}_{B_1}$, along with some finite contributions that must be properly identified and extracted.

The differential decay rate with bremsstrahlung RC can now be constructed out of $\mathsf{M}_B$ again by following a standard procedure, namely, by squaring it, summing over the final spins and over the photon polarization.

Thus, after a few algebraic manipulations, one gets
\begin{equation}
\sum_{\epsilon,s} |\mathsf{M}_B|^2 = \frac{e^2G^2}{2} \frac{8M_1^2}{mm_\nu} (b_1 + b_2 + b_3), \label{eq:mbf}
\end{equation}
where $b_1 \propto |\mathsf{M}_{B_1}|^2$, $b_2 \propto |\mathsf{M}_{B_2}|^2$, and $b_3$ contains the interference terms of the various $\mathsf{M}_{B_i}$ and also $|\mathsf{M}_{B_3}|^2$. Specifically, the former can be split into two terms, namely,
\begin{equation}
b_1 = b_1^{\mathrm{ir}} + b_1^{\mathrm{ic}}, \label{eq:b1f}
\end{equation}
where $b_1^{\mathrm{ir}}$ contains the infrared divergence,
\begin{equation}
b_1^{\mathrm{ir}} = \sum_{\epsilon} \left[\frac{l\cdot\epsilon}{l\cdot k} - \frac{p_2\cdot\epsilon}{p_2 \cdot k} \right]^2 \frac{M_1^2}{8} A_0,
\end{equation}
and the pending sum over the photon polarization in the above equation should be dealt with according to Coester's rule \cite{coester} to account for the longitudinal degree of polarization of the photon. At this point, notice that $\lambda^2=\omega^2-k^2$, where $\lambda$ is a fictitious mass given to the photon to regularize the infrared divergence. The additional term, $b_1^{\mathrm{ic}}$, does not contain any infrared divergence and is given by
\begin{equation}
b_1^{\mathrm{ic}} = b_{11} |f_+(q^2)|^2 + b_{12} \mathrm{Re} [f_+(q^2)f_-^*(q^2)] + b_{13} |f_-(q^2)|^2,
\end{equation}
where 
\begin{equation}
b_{11} = -\sum_{\epsilon} \left[\frac{l\cdot\epsilon}{l\cdot k}-\frac{p_2\cdot\epsilon}{p_2\cdot k}\right]^2 \omega \left[ 2E-\left[D+E(1-\beta\hat{\mathbf{l}}\cdot\mathbf{\hat{k}})\right] \left[1-\frac{m^2}{4M_1^2}\right] - \frac{m^2}{M_1} \right], \label{eq:b11}
\end{equation}
\begin{equation}
b_{12} = -\sum_{\epsilon} \left[\frac{l\cdot\epsilon}{l\cdot k}-\frac{p_2\cdot\epsilon}{p_2\cdot k}\right]^2 \frac{\omega m^2}{M_1} \left[1-\frac{D+E(1-\beta\hat{\mathbf{l}}\cdot\mathbf{\hat{k}})}{2M_1}\right], \label{eq:b12}
\end{equation}
and
\begin{equation}
b_{13} = -\sum_{\epsilon} \left[\frac{l\cdot\epsilon}{l\cdot k}-\frac{p_2\cdot\epsilon}{p_2\cdot k}\right]^2 \frac{\omega m^2}{4M_1^2} \left[D+E(1-\beta\hat{\mathbf{l}}\cdot\mathbf{\hat{k}})\right], \label{eq:b13}
\end{equation}
where
\begin{equation}
D = E_\nu^0 + (\mathbf{p}_2 + \mathbf{l}) \cdot \mathbf{\hat{k}},
\end{equation}
and 
\begin{equation}
\beta = \frac{l}{E}. \label{eq:beta}
\end{equation}
In Eqs.~(\ref{eq:b11})--(\ref{eq:b13}) the ordinary sum over the photon polarization can be safely used.

Now the terms $b_2$ and $b_3$ are also infrared convergent. They read
\begin{equation}
b_2 = b_{21} |f_+(q^2)|^2 + b_{22} \mathrm{Re} [f_+(q^2)f_-^*(q^2)] + b_{23} |f_-(q^2)|^2, \label{eq:b2f}
\end{equation}
with
\begin{equation}
b_{21} = \frac{1}{E(1-\beta\hat{\mathbf{l}}\cdot\mathbf{\hat{k}})} \left[2E_\nu^0- 2 \omega - D + \frac{m^2}{4M_1^2} D \right],
\end{equation}
\begin{equation}
b_{22} = -\frac{m^2}{2M_1^2} \frac{D}{E(1-\beta\hat{\mathbf{l}}\cdot\mathbf{\hat{k}})},
\end{equation}
and
\begin{equation}
b_{23}=-\frac{1}{2}b_{22}.
\end{equation}

Similarly,
\begin{equation}
b_3 = b_{31} |f_+(q^2)|^2 + b_{32} \mathrm{Re} [f_+(q^2)f_-^*(q^2)] + b_{33} |f_-(q^2)|^2, \label{eq:b3f}
\end{equation}
where
\begin{eqnarray}
b_{31} & = & -\left[ \frac{E}{\omega} \frac{1-\beta^2}{1-\beta\hat{\mathbf{l}} \cdot\hat{\mathbf{k}}} - \frac{EE_2-\mathbf{p}_2 \cdot \mathbf{l}}{E_2\omega - \mathbf{p}_2 \cdot \mathbf{k}} \right] \left[\frac{2E_\nu}{E(1-\beta\hat{\mathbf{l}}\cdot\hat{\mathbf{k}})} + \left[\frac{D}{E(1-\beta\hat{\mathbf{l}}\cdot\hat{\mathbf{k}})} + 1 \right] \left[-1+\frac{m^2}{4M_1^2}\right] \right] \nonumber \\
&  & \mbox{} + \frac{2E_\nu}{\omega} \left[ \frac{1}{1-\beta\hat{\mathbf{l}} \cdot \hat{\mathbf{k}}} - \frac{E_2}{E_2-\mathbf{p}_2 \cdot \hat{\mathbf{k}}} \right] + \left[\frac{M_1-E_2+\beta\mathbf{p}_2\cdot \hat{\mathbf{l}}}{\omega(1-\beta\hat{\mathbf{l}}\cdot\hat{\mathbf{k}})} - \frac{M_1E_2-M_2^2}{E_2\omega - \mathbf{p}_2 \cdot \mathbf{k}} \right] \left[-1+\frac{m^2}{4M_1^2}\right],
\end{eqnarray}
\begin{equation}
b_{32} = \frac{m^2}{2M_1^2} \left[ \left[\frac{E}{\omega}\frac{1-\beta^2}{1-\beta\hat{\mathbf{l}}\cdot\hat{\mathbf{k}}}-\frac{EE_2-{\mathbf{p}_2}\cdot{\mathbf{l}}}{E_2\omega-{\mathbf{p}_2}\cdot{\mathbf{k}}}\right]\left[\frac{D}{E(1-
\beta\hat{\mathbf{l}}\cdot\hat{\mathbf{k}})}+1 \right ]-\left[\frac{M_1-E_2+\beta{\mathbf{p}_2}\cdot{\hat{\mathbf{l}}}} {\omega(1-\beta\hat{\mathbf{l}}\cdot\hat{\mathbf{k}})}-\frac{M_1E_2-M_2^2}{E_2\omega-\mathbf{p}_2\cdot\mathbf{k}}\right ]\right],
\end{equation}
and
\begin{equation}
b_{33} = - \frac12 b_{32}.
\end{equation}

The differential decay rate with bremsstrahlung RC can be constructed out of Eq.~(\ref{eq:mbf}) as
\begin{equation}
d\Gamma_B = \sum_{i=1}^{3} d\Gamma_{B_i},
\end{equation}
where the different contributions $d\Gamma_{B_i}$ are proportional to the corresponding $b_i$ factors defined in Eqs.~(\ref{eq:b1f}), (\ref{eq:b2f}), and (\ref{eq:b3f}). For definiteness,
\begin{equation}
d\Gamma_{B_1} = d\Gamma_{B_1}^{\mathrm{ir}} + d\Gamma_{B_1}^{\mathrm{ic}}, \label{eq:dg1}
\end{equation}
where the first summand in Eq.~(\ref{eq:dg1}) is the one that contains the infrared divergence. The procedure to deal with this kind of contribution has been described in detail in Ref.~\cite{jl2011}. Without further ado, the resultant expression reads
\begin{equation}
d\Gamma_{B_1}^{\mathrm{ir}}=\frac{\alpha}{\pi}d\Omega A_0I_{0n}(E,E_2),
\end{equation}
where $A_0$ is defined in Eq.~(\ref{eq:a0}) and $I_{0n}$, originally given in Ref.~\cite{ginsKm3}, was ultimately corrected in Ref.~\cite{cir2}. The latter result will be borrowed here. Thus, the function $I_{0n}$, adapted to the current notation, reads
\begin{eqnarray}
I_{0n}(E,E_2) & = & 2\log \frac{m}{\lambda} \left[ \frac{1}{\beta^\prime} \mathrm{arctanh}\, \beta^\prime - 1 \right] + \frac{1}{\beta^\prime} \mathrm{arctanh}\, \beta^\prime \log \frac{\Delta}{m^2} - \log \frac{M_2}{m} - (\mathrm{arctanh}\, \beta^\prime)^2 \nonumber \\
&  & \mbox{} + \left( \mathrm{arctanh}\, \beta + \mathrm{arctanh}\, \beta_2 \right)^2 + \frac{1}{\beta^\prime} \mathrm{arctanh}\, \beta^\prime \log \frac{2\beta^\prime \chi (l\cdot p_2)^2(1-\tau_0^2)^2}{M_1^2(E_m-E)(W_2-E_2)} \nonumber \\
&  & \mbox{} + \log \frac{4M_1^2(E_m-E)(W_2-E_2)}{\eta_m^2} + \frac{1}{2\beta^\prime} \left[ L(\eta_1) - L(1/\eta_1) + L(\eta_2) - L(1/\eta_2)\right] \nonumber \\
&  & \mbox{} + \frac{2}{\beta^\prime} \left[ \log \tau_m \log \frac{1-\tau_0 \tau_m}{1-\tau_m/\tau_0} - L(\tau_0 \tau_m) + L(\tau_m/\tau_0) + L(\tau_0^2) + \frac{\pi^2}{6} \right], \label{eq:i0n}
\end{eqnarray}
where $\beta^\prime$ is defined in Eq.~(\ref{eq:betap}), and
\begin{equation}
\beta_2 =  \frac{p_2}{E_2} ,\label{eq:beta2}
\end{equation}
\begin{equation}
\tau_0 = \sqrt{\frac{1-\beta^\prime}{1+\beta^\prime}},
\end{equation}
\begin{equation}
\tau_m = \frac{(E-l)(E_2-p_2)}{mM_2},
\end{equation}
\begin{equation}
\chi = \frac{\Delta^2}{2a} \frac{(E_m-E)(W_2-E_2)}{a(E_m-E)(W_2-E_2) - m^2(E_m-E)^2 - M_2^2(W_2-E_2)^2},
\end{equation}
\begin{equation}
\eta_{1,2} = \frac{1-2\chi \pm \sqrt{{\beta^\prime}^2 + 4\chi^2 - 4\chi}}{1+\beta^{\prime }}.
\end{equation}

The quantities $W_2$, $\eta_m$, $a$, and $\Delta$ read
\begin{equation}
W_2 = \frac{M_1^2 + M_2^2 - m^2}{2M_1},
\end{equation}
\begin{equation}
a = 2p_2 \cdot l =2(EE_2 - p_2 l y_0), \label{eq:aa}
\end{equation}
\begin{equation}
\Delta = \sqrt{a^2 - 4m^2M_2^2} =\beta^\prime a,
\end{equation}
where
\begin{equation}
\eta_m = 2p_2 l(1+y_0),
\end{equation}
with
\begin{equation}
y_0 = \frac{{E_\nu^0}^2 - p_2^2 -l^2}{2p_2l}.
\end{equation}

On the other hand, the infrared-convergent piece, expressed as an integral over the kinematical variables, is
\begin{equation}
d\Gamma_{B_1}^{\mathrm{ic}} =\frac{\alpha}{\pi} d\Omega [\Lambda_{1n} |f_+(q^2)|^2 + \Lambda_{2n} \mathrm{Re}\, [f_+(^2)f_-^*(q^2)] + \Lambda_{3n} |f_-(q^2)|^2],
\end{equation}
where
\begin{equation}
\Lambda_{1n,2n,3n} = \frac{p_2l}{4\pi} \frac{8}{M_1^2} \int_{-1}^1 dx \int_{-1}^{y_0} dy  \int_0^{2\pi} d\phi_k \frac{\omega}{D} [b_{11},b_{12},b_{13}], \label{eq:lb1}
\end{equation}
and the integral form of the $\Lambda_{kn}$ functions follows the choice of orientation of the coordinate axes, namely, $\ell^+$ is emitted along the $+z$ axis and $\pi^-$ is emitted in the first or fourth quadrant of the $(x,z)$ plane \cite{jl2011}. Thus, $x=\hat{\mathbf{k}} \cdot \hat{\mathbf{l}}$, $y=\hat{\mathbf{p}}_2 \cdot \hat{\mathbf{l}}$, and $\phi_{k}$ is the azimuthal angle of the momentum of the photon.

Similarly, $d\Gamma_{B_2}$ and $d\Gamma_{B_3}$, which are also infrared convergent, are given by
\begin{equation}
d\Gamma_{B_2} = \frac{\alpha}{\pi} d\Omega [\Lambda_{4n} |f_+(q^2)|^2 + \Lambda_{5n} \mathrm{Re} [f_+(q^2)f_-^*(q^2)] + \Lambda_{6n} |f_-(q^2)|^2], \label{eq:b2}
\end{equation}
and
\begin{equation}
d\Gamma_{B_3} = \frac{\alpha}{\pi} d\Omega [\Lambda_{7n} |f_+(q^2)|^2 + \Lambda_{8n} \mathrm{Re} [f_+(q^2)f_-^*(q^2)] + \Lambda_{9n} |f_-(q^2)|^2], \label{eq:b3}
\end{equation}
where
\begin{equation}
\Lambda_{4n,5n,6n} = \frac{p_2l}{4\pi} \frac{8}{M_1^2} \int_{-1}^1 dx \int_{-1}^{y_0} dy \int_0^{2\pi} d\phi_k \frac{\omega}{D} [b_{21},b_{22},b_{23}], \label{eq:lb2}
\end{equation}
and
\begin{equation}
\Lambda_{7n,8n,9n} = \frac{p_2l}{4\pi} \frac{8}{M_1^2} \int_{-1}^1 dx \int_{-1}^{y_0} dy \int_0^{2\pi} d\phi_k \frac{\omega}{D} [b_{31},b_{32},b_{33}]. \label{eq:lb3}
\end{equation}

Gathering together partial results, the differential decay rate $d\Gamma_B$ can be expressed in a compact form as
\begin{equation}
d\Gamma_B = \frac{\alpha}{\pi} d\Omega [A_0 I_{0n} + A^\prime_{Bn}], \label{eq:dgb}
\end{equation}
where
\begin{equation}
A_{Bn}^\prime = A_{1n}^{(B)} |f_+(q^2)|^2 + A_{2n}^{(B)} \mathrm{Re} [f_+(q^2)f_-^*(q^2)] + A_{3n}^{(B)} |f_-(q^2)|^2, \label{eq:abnp}
\end{equation}
with
\begin{subequations}
\begin{eqnarray}
A_{1n}^{(B)} & = & \Lambda_{1n} + \Lambda_{4n} + \Lambda_{7n}, \\
A_{2n}^{(B)} & = & \Lambda_{2n} + \Lambda_{5n} + \Lambda_{8n}, \\
A_{3n}^{(B)} & = & \Lambda_{3n} + \Lambda_{6n} + \Lambda_{9n}, 
\end{eqnarray}
\end{subequations}

At this point the first final result has been reached. The DP of $K_{l3}^0$ decays with radiative corrections to order $(\alpha/\pi)(q/M_1)$ is obtained by adding Eqs.~(\ref{eq:dgv}) and (\ref{eq:dgb}) to obtain $d\Gamma(K_{l3}^0)$. The integrations over the three-momentum of the real photon in Eqs.~(\ref{eq:lb1}) and (\ref{eq:lb2})--(\ref{eq:lb3}) can be performed numerically. It turns out that the remaining photon integrals can be performed analytically. This will be done in the next section. This way a completely analytical result will be obtained. This will be the second final result.

\subsection{Analytical integrations}

The triple integrals indicated in Eqs.~(\ref{eq:lb1}) and (\ref{eq:lb2})--(\ref{eq:lb3}) can in principle be computed analytically to meet the same standards as in Refs.~\cite{jl2011,jl2012}. However, the presence of the factor $1/p_2\cdot k$ in the sum over the photon polarization in all the infrared-convergent pieces, or equivalently the factor $\mathbf{p}_2 \cdot \mathbf{k}$ in the denominators of some functions $b_{ij}$, makes the calculation rather involved. There is one approximation that could be used, namely,
\begin{equation}
\frac{1}{p_2\cdot k} \approx \frac{1}{p_1\cdot k} + \frac{q\cdot k}{(p_1\cdot k)^2},
\end{equation}
provided the momentum transfer is small. However, for $K_{l3}$ decays it is not the case, so the approximation is useless in the present analysis. There is however a symmetry property that can still be exploited: the transformation properties of the integrands under rotations. Thus, the right orientation of the coordinate axes will simplify the task enormously. Skipping details, the analytical form of the $\Lambda_{kn}$ functions reads
\begin{eqnarray}
\frac{M_1^2}{4p_2l} \Lambda_{1n} & = & \left[2E-\frac{m^2}{M_1}\right] \left[(1-\beta^2)\theta_2 + \frac{M_2^2}{E_2^2} \theta_2^\prime \right] - \left[\frac{2E}{M_1}-\frac{m^2}{M_1^2} \right]\left[2E_2\theta_3+2E\theta_3^\prime-\frac{2}{E}\zeta_{11}-\frac{2}{E_2}\zeta_{11}^\prime + 4J_{1n} \right] \nonumber \\
&  & \mbox{} - \left[\frac{4M_1^2}{m^2}-1 \right] \frac{M_1^2}{4p_2l} \Lambda_{3n}, \label{eq:lb1a}
\end{eqnarray}
\begin{equation}
\frac{M_1^2}{4p_2l} \Lambda_{2n} = \frac{m^2}{M_1^2}\left[-\frac12 \eta_0+M_1(1-\beta^2)\theta_2+\frac{M_1M_2^2}{2E_2^2}\theta_2^\prime-\left[\frac{E}{2}(1-\beta^2)+2E_2\right]\theta_3-\left[E-\frac{M_2^2}{2E_2}\right]\theta_3^\prime+\frac{2}{E}\zeta_{11}+\frac{\zeta_{11}^\prime}{E_2}-2J_{1n} \right],
\end{equation}
\begin{equation}
\frac{M_1^2}{4p_2l} \Lambda_{3n} = \frac{m^2}{4M_1^2}\left[2\eta_0+E(1-\beta^2)\theta_3+\frac{M_2^2}{E_2^2}(M_1\theta_2^\prime-E_2\theta_3^\prime) - 2E\theta_3^\prime+\frac{2}{E_2}\zeta_{11}^\prime-4J_{1n}\right],
\end{equation}
\begin{equation}
\frac{M_1^2}{4p_2l} \Lambda_{4n} = \left[-2E_\nu^0-\left[1-\frac{m^2}{4M_1^2}\right]\beta p_2y_0-3E+\beta l\right]\theta_3+(2E_\nu^0+3E)\theta_4 + 3l\theta_5+\frac{E_{\nu}^0}{E}\theta_{7}-\frac{1}{2E}\theta_{9}+\left[1-\frac{m^2}{4M_1^2}\right]\frac{\zeta_{11}}{E},
\end{equation}
\begin{equation}
\frac{M_1^2}{4p_2l} \Lambda_{5n} = -\frac{m^2}{2M_1^2} \left[\beta p_2y_0\theta_3-\frac{1}{E}\zeta_{11}\right],
\end{equation}
\begin{equation}
\Lambda_{6n} = -\frac12 \Lambda_{5n},
\end{equation}
\begin{eqnarray}
\frac{M_1^2}{4p_2l} \Lambda_{7n} & = & \frac{M_1^2}{2p_2l} \left[\frac{M_1^2}{m^2}-\frac14 \right] \Lambda_{8n} - 2(1-\beta^2) \left[E_\nu^0\theta_2 - E(\theta_3-\theta_2) - \frac{\theta_6}{2}\right]  + 2E_\nu^0 (\theta_3-\theta_3^\prime) \nonumber \\
&  & \mbox{} - 2\left[E(\theta_4-\theta_3) - E_2(\theta_4^\prime-\theta_3^\prime) + \frac{\theta_7}{2} - \frac{\theta_7^\prime}{2} \right] + \frac{2E_\nu^0}{M_1} \left[E_2\theta_3 + E\theta_3^\prime -\frac{\zeta_{11}}{E} - \frac{\zeta_{11}^\prime}{E_2} + 2J_{1n}\right] \nonumber \\
&  & \mbox{} - \frac{2EE_2}{M_1} \left[\theta_4-\theta_3+\theta_4^\prime -\theta_3^\prime +\frac{\theta_7}{2E}+\frac{\theta_7^\prime}{2E_2}\right] + \frac{1}{M_1} \left[\frac{\zeta_{21}}{E}+\frac{\zeta_{21}^\prime }{E_2} \right] \nonumber \\
&  & \mbox{} - \frac{2(EE_2-p_2ly_0)}{M_1^2} \left[\frac{p_2ly_0\theta_3-\zeta_{11}}{E} - \frac{2p_2ly_0\theta_3^\prime -\zeta_{11}^\prime}{E_2} \right] - \frac{1}{2M_1^2} \left[\frac{I_n}{E}+\frac{I_n^\prime}{E_2} \right] - \frac{4mM_2}{p_2l}(J_{2n}+J_{3n}),
\end{eqnarray}
and
\begin{equation}
\frac{M_1^2}{4p_2l} \Lambda_{8n} = \frac{m^2}{2M_1^2} \left[ 2\eta_0 +\left[E(1-\beta^2)+E_2-M_1\right]\theta_3 + \left[M_1-E-\frac{M_2^2}{E_2}\right]\theta_3^\prime + \frac{\zeta_{11}^\prime}{E_2}-\frac{\zeta_{11}}{E}-2J_{1n}\right], 
\end{equation}
\begin{equation}
\Lambda_{9n} = -\frac12 \Lambda_{8n}. \label{eq:lb9a}
\end{equation}
The $\theta_m$, $\eta_0$, and $\zeta_{ij}$ functions can be found in Ref.~\cite{rfm04}. The $\theta_m^\prime$ and $\zeta_{ij}^\prime$ functions are obtained by making the replacements $p_2 \leftrightarrow l$ in the corresponding $\theta_m$ and $\zeta_{ij}$ functions. The additional functions involved in the above expressions are
\begin{equation}
J_{1n} = \frac{1}{\beta\beta_2} \left[-(1+\beta\beta_2) + (\beta+\beta_2)(\mathrm{arctanh}\, \beta+\mathrm{arctanh}\, \beta_2) - (1-\beta\beta_2y_0)(-1+\beta^\prime \mathrm{arctanh}\, \beta^\prime)\right].
\end{equation}
\begin{eqnarray}
\frac{I_n}{4p_2^2l^2} & = & \frac{E_\nu^0}{p_2^2}\eta_0 + \frac{\beta E_\nu^0+l-p_2}{2\beta p_2^2}\theta_0 - \frac{EE_\nu^0}{p_2^2}(\theta_3-\theta_4) +\frac{1}{2p_2^2\beta^2}\left[3{E_\nu^0}^2-l^2+3E(E+2E_\nu^0)\right] (\theta_3-\theta_4-\beta\theta_5) \nonumber \\
&  & \mbox{} + \left[ y_0^2 - \frac{{E_\nu^0}^2}{2p_2^2} \right] \theta_3 - \frac{3E}{2p_2^2}(E_\nu^0+E)\theta_{10} - \frac{3E}{2p_2l}(E+E_\nu^0)(\theta_{12} - \theta_{13}) + \frac{y_0}{2} \theta_{12} + \frac{3E}{2p_2} Y_1 + \frac{1}{2\beta^2}Y_3 - \frac{2y_0}{p_2l} \zeta_{11}, \nonumber \\
\end{eqnarray}
\begin{eqnarray}
\frac{8M_1^2}{mM_2} J_{2n} & = & \frac{1+{\beta^\prime}^2}{1-{\beta^\prime}^2} - 2\mathrm{arccosh} \left[\frac{1}{\sqrt{1-{\beta^\prime}^2}} \right] \left[\mathrm{arccosh} \left[\frac{1}{\sqrt{1-{\beta^\prime}^2}} \right] + \frac{2\beta^\prime}{1-{\beta^\prime}^2} \right] -\frac{2(1+\beta\beta_2)^2}{(1-\beta^2)(1-{\beta_2}^2)} + 1 \nonumber \\
&  & \mbox{} + 2\mathrm{arccosh} \left[\frac{1+\beta\beta_2}{\sqrt{1-\beta^2}\sqrt{1-\beta_2^2}}\right] \left[\mathrm{arccosh} \left[\frac{1+\beta\beta_2}{\sqrt{1-\beta^2}\sqrt{1-\beta_2 ^2}}\right] + \frac{2(\beta+\beta_2)(1+\beta\beta_2)}{(1-\beta^2)(1-\beta_2^2)}\right] ,
\end{eqnarray}
and
\begin{equation}
\frac{2M_1^2}{a} J_{3n} = \frac{p_2l}{mM_2}\eta_0 + \frac{a\beta^\prime}{2mM_2} \mathrm{arccosh} \left[ \frac{a}{2mM_2}\right] - \frac{EE_2}{mM_2} (\beta+\beta_2) \mathrm{arccosh} \left[\frac{1+\beta\beta_2}{\sqrt{1-\beta^2}\sqrt{1-\beta_2^2}} \right],
\end{equation}
where the factors $\beta$, $\beta^\prime$, $\beta_2$, and $a$ are defined in Eqs.~(\ref{eq:beta}), (\ref{eq:betap}), (\ref{eq:beta2}), and (\ref{eq:aa}), respectively.

\section{\label{sec:finr}Final results and discussions}

The differential decay rate of $K_{l3}^0$ decays in the variables $E$ and $E_2$ (that is, the DP) including radiative corrections to order $(\alpha/\pi)(q/M_1)$, is given by
\begin{equation}
d\Gamma(K_{l3}^0) = d\Gamma_V + d\Gamma_B. \label{eq:dkl3t}
\end{equation}
$d\Gamma_V$ is given by Eq.~(\ref{eq:dgv}). For $d\Gamma_B$ two forms are available. The first one contains the triple integration over the real photon variables standing so it can be performed numerically. It is given by 
Eq.~(\ref{eq:dgb}), which is expressed in terms of the functions $\Lambda_{kn}$ introduced in Eqs.~(\ref{eq:lb1}) and (\ref{eq:lb2})--(\ref{eq:lb3}). The infrared divergence and the finite terms that come along with it have been explicitly and analytically extracted, however; the infrared divergence is of course canceled out in the sum in (\ref{eq:dkl3t}). The second form of $d\Gamma_B$ is completely analytical; the integration over the photon variables has been explicitly computed and the analytical versions of the functions $\Lambda_{kn}$ are thus given in Eqs.~(\ref{eq:lb1a})--(\ref{eq:lb9a}).

The main result obtained here can be cast into the compact form
\begin{equation}
d\Gamma(K_{l3}^0) = \frac{G_F^2}{32\pi^3} |V_{us}|^2 M_1^3 \left[A_0^\prime + \frac{\alpha}{\pi}A_n^\prime \right] dEdE_2. \label{eq:dgfin}
\end{equation}
$A_0^\prime$ has been previously computed; it is given in Eq.~(16) of Ref.~\cite{jl2011}. On the other hand, $A_n^\prime$ can be written as
\begin{eqnarray}
A_n^\prime & = & A_0^\prime (\mathrm{Re}\, \Phi_{n} + I_{0n}) + A_{Vn}^\prime \mathrm{Re}\, \Phi_n^\prime + A_{Bn} \nonumber \\
& = & A_{10} |f_+^\prime(q^2,p_+\cdot l)|^2 + A_{20} \mathrm{Re} [f_+^\prime(q^2,p_+\cdot l) {f_-^\prime}^*(q^2,p_+\cdot l)] + A_{30} |f_-^\prime(q^2,p_+\cdot l)|^2, \label{eq:anp}
\end{eqnarray}
where $\Phi_{1n}$, $\Phi_n^\prime$, and $I_{0n}$ have also been previously computed; the first two are given in Ref.~\cite{tun93} and the third one is given in Ref.~\cite{cir2}. They are nevertheless listed in this paper in Eqs.~(\ref{eq:phin})--(\ref{eq:phinp}) and (\ref{eq:i0n}), respectively, for the sake of completeness. The new expression $A_{Bn}$ is thus the main contribution. It is defined in Eq.~(\ref{eq:abnp}) and is written in terms of the $\Lambda_{kn}$ functions discussed above.

Equation (\ref{eq:dgfin}) has some advantages: it contains all the terms of the order of $(\alpha/\pi)(q/M_1)$, does not have an infrared divergence, does not contain an ultraviolet cutoff, and is not compromised by any model dependence of RC. Despite its length, it is basically simple and organized in a way that is easy to handle. A common practice advocated in experimental setups is the implementation of kinematical cuts to the observed electron and emitted kaon kinematical variables. As a result, only a region of points and not the full DP is accessible in an experiment. However, on each one point of the DP the photon momentum integration limits do depend on the values of $(E,l)$ and $(E_2 ,p_2)$ of that point. Thus, the common kinematical cuts are automatically taken into account in the integration limits of the emitted photons at each point. Therefore, the main usefulness of the analytical result lies in that it can be incorporated into a Monte Carlo simulation of an experimental analysis, with a considerable reduction of the computational effort required by the triple integration pending in the first form of the result.

In order to ensure the reliability of the results presented here, they have been cross-checked by performing numerically the triple integrals involved in Eqs.~(\ref{eq:lb1}) and (\ref{eq:lb2})--(\ref{eq:lb3}) and then comparing these results with their analytical counterparts in Eqs.~(\ref{eq:lb1a})--(\ref{eq:lb9a}). The agreement found is very good. A further comparison, at least partially, can be performed with other calculations already published. The closest results are those presented in Table II of Ref.~\cite{cir2}, which corresponds to the RC to the differential decay rate of the $K_{e3}^0$ mode. These results can be contrasted with the ones obtained here for the same mode and listed in Table \ref{t:ecrA102030}. The agreement in practically the totality of the kinematical region is remarkable. For completeness, the corresponding results for the $K_{\mu 3}^0$ mode are presented in Table \ref{t:mcrA102030}. As expected, in this case the contributions emerging from the $f_+f_-$ and $f_-^2$ parts are non-negligible compared to the leading $f_+^2$ one.

To close this paper, it should be pointed out that the expressions obtained here are very useful for processes where the momentum transfer is not small so that it cannot be neglected. Thus, they are suitable to any $M_{l3}^0$ decay, whether $M$ be $\pi^0$, $K^0$, $D^0$, or even $B^0$. An estimated upper bound to the theoretical uncertainty of 1.2\% can be made \cite{jl2011} so this should be acceptable with an experimental precision of 2\%--3\%. It should be emphasized, however, that the restriction imposed here that bremsstrahlung photons be experimentally discriminated either by direct detection or indirectly by energy-momentum conservation limits the scope of the results to the TBR of the DP. Further reduction of the theoretical uncertainty would require the relaxation of this restriction, which falls into the realm of the so-called four-body region of the DP. This calculation, however, requires a non-negligible extra effort that will be attempted in the near future.

\begingroup
\squeezetable
\begin{table}
\caption{\label{t:ecrA102030}Radiative correction $(\alpha/\pi)A_n^\prime$, Eq.~(\ref{eq:anp}), in the TBR of the process $K^0\to \pi^-+e^++\nu_e$. The entries correspond to $(\alpha/\pi)A_{10} \times 10$; $(\alpha/\pi)A_{20}$  and $(\alpha/\pi)A_{30}$ are negligible for this mode. The energies $E$ and $E_2$ are given in $\textrm{GeV}$.}
\begin{ruledtabular}
\begin{tabular}{crrrrrrrrr}
$E_2\backslash E$ & $ 0.0123$ & $ 0.0370$ & $ 0.0617$ & $ 0.0864$ & $ 0.1111$ & $ 0.1358$ & $ 0.1604$ & $ 0.1851$ & $ 0.2098$ \\
\hline
$0.2592$ & $ 0.1736$ & $ 0.2138$ & $ 0.1606$ & $ 0.0675$ & $-0.0401$ & $-0.1425$ & $-0.2200$ & $-0.2491$ & $-0.1896$ \\
$0.2468$ &           & $ 0.2289$ & $ 0.2145$ & $ 0.1422$ & $ 0.0428$ & $-0.0616$ & $-0.1501$ & $-0.1973$ & $-0.1602$ \\
$0.2345$ &           & $ 0.1870$ & $ 0.2152$ & $ 0.1630$ & $ 0.0745$ & $-0.0255$ & $-0.1149$ & $-0.1673$ & $-0.1383$ \\
$0.2222$ &           &           & $ 0.1946$ & $ 0.1664$ & $ 0.0911$ & $-0.0019$ & $-0.0889$ & $-0.1428$ & $-0.1178$ \\
$0.2098$ &           &           & $ 0.1495$ & $ 0.1588$ & $ 0.0993$ & $ 0.0151$ & $-0.0676$ & $-0.1209$ & $-0.0977$ \\
$0.1975$ &           &           &           & $ 0.1399$ & $ 0.1011$ & $ 0.0278$ & $-0.0492$ & $-0.1005$ & $-0.0774$ \\
$0.1851$ &           &           &           &           & $ 0.0966$ & $ 0.0368$ & $-0.0330$ & $-0.0811$ & $-0.0563$ \\
$0.1728$ &           &           &           &           & $ 0.0834$ & $ 0.0422$ & $-0.0187$ & $-0.0623$ & $-0.0331$ \\
$0.1604$ &           &           &           &           &           & $ 0.0430$ & $-0.0062$ & $-0.0439$ & $-0.0200$ \\
$0.1481$ &           &           &           &           &           &           & $ 0.0042$ & $-0.0251$ &           \\
\end{tabular}
\end{ruledtabular}
\end{table}
\endgroup

\begingroup
\squeezetable
\begin{table}
\caption{\label{t:mcrA102030}Radiative correction $(\alpha/\pi)A_n^\prime$, Eq.~(\ref{eq:anp}), in the TBR of the process $K^0\to \pi^-+\mu^++\nu_\mu$. The entries correspond to (a) $(\alpha/\pi)A_{10}\times 10$, (b) $(\alpha/\pi)A_{20}\times 10^2$, and (c) $(\alpha/\pi)A_{30}\times 10^3$. The energies $E$ and $E_2$ are given in $\textrm{GeV}$.}
\begin{ruledtabular}
\begin{tabular}{lrrrrrrrrr}
$E_2\backslash E$ & $ 0.1131$ & $ 0.1280$ & $ 0.1429$ & $ 0.1578$ & $ 0.1727$ & $ 0.1876$ & $ 0.2025$ & $ 0.2174$ & $ 0.2322$ \\
\hline
(a)      &           &           &           &           &           &           &           &           &           \\
$0.2480$ &           &           &           & $-0.0405$ & $-0.0228$ & $-0.0199$ & $-0.0207$ & $-0.0207$ & $-0.0144$ \\
$0.2361$ &           & $ 0.0513$ & $ 0.0490$ & $ 0.0398$ & $ 0.0272$ & $ 0.0134$ & $ 0.0001$ & $-0.0101$ & $-0.0110$ \\
$0.2242$ & $ 0.0846$ & $ 0.0808$ & $ 0.0709$ & $ 0.0573$ & $ 0.0412$ & $ 0.0239$ & $ 0.0073$ & $-0.0058$ & $-0.0088$ \\
$0.2123$ & $ 0.1005$ & $ 0.0890$ & $ 0.0775$ & $ 0.0632$ & $ 0.0464$ & $ 0.0283$ & $ 0.0107$ & $-0.0035$ & $-0.0069$ \\
$0.2004$ & $ 0.1384$ & $ 0.0910$ & $ 0.0777$ & $ 0.0637$ & $ 0.0474$ & $ 0.0296$ & $ 0.0121$ & $-0.0021$ & $-0.0054$ \\
$0.1885$ &           & $ 0.1009$ & $ 0.0748$ & $ 0.0610$ & $ 0.0458$ & $ 0.0290$ & $ 0.0123$ & $-0.0013$ &           \\
$0.1766$ &           &           & $ 0.0742$ & $ 0.0562$ & $ 0.0422$ & $ 0.0270$ & $ 0.0115$ & $-0.0011$ &           \\
$0.1647$ &           &           &           & $ 0.0513$ & $ 0.0370$ & $ 0.0236$ & $ 0.0099$ & $-0.0015$ &           \\
$0.1528$ &           &           &           &           & $ 0.0308$ & $ 0.0190$ & $ 0.0073$ &           &           \\
$0.1409$ &           &           &           &           &           & $ 0.0126$ & $-0.0006$ &           &           \\
(b)      &           &           &           &           &           &           &           &           &           \\
$0.2480$ &           &           &           & $-0.0381$ & $-0.0227$ & $-0.0198$ & $-0.0194$ & $-0.0186$ & $-0.0153$ \\
$0.2361$ &           & $ 0.0575$ & $ 0.0428$ & $ 0.0267$ & $ 0.0124$ & $ 0.0003$ & $-0.0095$ & $-0.0164$ & $-0.0207$ \\
$0.2242$ & $ 0.1494$ & $ 0.1072$ & $ 0.0746$ & $ 0.0487$ & $ 0.0275$ & $ 0.0099$ & $-0.0045$ & $-0.0157$ & $-0.0269$ \\
$0.2123$ & $ 0.2290$ & $ 0.1475$ & $ 0.1007$ & $ 0.0669$ & $ 0.0402$ & $ 0.0184$ & $ 0.0003$ & $-0.0148$ & $-0.0358$ \\
$0.2004$ & $ 0.4634$ & $ 0.2020$ & $ 0.1304$ & $ 0.0863$ & $ 0.0536$ & $ 0.0274$ & $ 0.0057$ & $-0.0135$ & $-0.0576$ \\
$0.1885$ &           & $ 0.3368$ & $ 0.1730$ & $ 0.1103$ & $ 0.0693$ & $ 0.0380$ & $ 0.0121$ & $-0.0121$ &           \\
$0.1766$ &           &           & $ 0.2635$ & $ 0.1449$ & $ 0.0893$ & $ 0.0508$ & $ 0.0197$ & $-0.0114$ &           \\
$0.1647$ &           &           &           & $ 0.2107$ & $ 0.1177$ & $ 0.0671$ & $ 0.0285$ & $-0.0145$ &           \\
$0.1528$ &           &           &           &           & $ 0.1674$ & $ 0.0895$ & $ 0.0380$ &           &           \\
$0.1409$ &           &           &           &           &           & $ 0.1218$ & $-0.0077$ &           &           \\
(c)      &           &           &           &           &           &           &           &           &           \\
$0.2480$ &           &           &           & $-0.0135$ & $-0.0145$ & $-0.0177$ & $-0.0213$ & $-0.0254$ & $-0.0325$ \\
$0.2361$ &           & $ 0.0155$ & $ 0.0059$ & $-0.0034$ & $-0.0119$ & $-0.0201$ & $-0.0284$ & $-0.0387$ & $-0.0615$ \\
$0.2242$ & $ 0.0767$ & $ 0.0509$ & $ 0.0314$ & $ 0.0154$ & $ 0.0015$ & $-0.0117$ & $-0.0257$ & $-0.0441$ & $-0.0911$ \\
$0.2123$ & $ 0.1610$ & $ 0.1033$ & $ 0.0699$ & $ 0.0451$ & $ 0.0243$ & $ 0.0047$ & $-0.0162$ & $-0.0448$ & $-0.1312$ \\
$0.2004$ & $ 0.4178$ & $ 0.1883$ & $ 0.1258$ & $ 0.0869$ & $ 0.0564$ & $ 0.0287$ & $-0.0009$ & $-0.0424$ & $-0.2234$ \\
$0.1885$ &           & $ 0.3884$ & $ 0.2114$ & $ 0.1447$ & $ 0.0994$ & $ 0.0606$ & $ 0.0200$ & $-0.0384$ &           \\
$0.1766$ &           &           & $ 0.3832$ & $ 0.2294$ & $ 0.1570$ & $ 0.1019$ & $ 0.0466$ & $-0.0360$ &           \\
$0.1647$ &           &           &           & $ 0.3824$ & $ 0.2386$ & $ 0.1557$ & $ 0.0788$ & $-0.0468$ &           \\
$0.1528$ &           &           &           &           & $ 0.3751$ & $ 0.2284$ & $ 0.1138$ &           &           \\
$0.1409$ &           &           &           &           &           & $ 0.3285$ & $-0.0231$ &           &           \\
\end{tabular}
\end{ruledtabular}
\end{table}
\endgroup

\begin{acknowledgments}
The authors are grateful to Consejo Nacional de Ciencia y Tecnolog{\'\i}a (Mexico) for partial support. J.J.T.\ and A.M.\ were partially supported by Comisi\'on de Operaci\'on y Fomento de Actividades Acad\'emicas (Instituto
Polit\'ecnico Nacional). R.F.-M.\ was also partially supported by Fondo de Apoyo a la Investigaci\'on (Universidad Aut\'onoma de San Luis Potos{\'\i}).
\end{acknowledgments}

\end{document}